\documentclass[twoside,   onecolumn,   aps,   10pt,  prl, 
  floatfix,
  showpacs,
  superscriptaddress,
]{revtex4-1}

\usepackage{graphicx}
\usepackage{dcolumn}
\usepackage{amsmath}
\usepackage{bm}
\usepackage{float}
\usepackage{color}
\usepackage[flushleft]{threeparttable}
\usepackage{tabulary}
\newcommand{\fig}[1]{Fig.~\ref{#1}}

\newcommand{\addbirla}{Department of Physics, Birla Institute of Technology and Science Pilani, K K Birla Goa Campus, Goa - 403 726, India}
\newcommand{\addmctwo}{Department of Microtechnology and Nanoscience, Chalmers University of Technology, SE-41296, Gothenburg, Sweden}
\newcommand{\addphysics}{Department of Physics, Chalmers University of Technology, SE-41296, Gothenburg, Sweden}
\newcommand{\addtaipei}{Institute of Physics, Academia Sinica, Taipei 10617, Taiwan}
\newcommand{\addtaipe}{Center for Condensed Matter Science, National Taiwan University,  Taipei 10617, Taiwan }

\begin{document}

\title{
    Resistivity Anomaly     in Weyl Semimetal candidate Molybdenum Telluride
}
\author{Dhavala Suri}
\affiliation{\addbirla}
\author{Christopher Linder\"{a}lv}
\affiliation{\addphysics}
\author{Bogdan Karpiak}
\author{Linnea Andersson}
\affiliation{\addmctwo}
\author{Sandeep Kumar Singh}
\affiliation{\addphysics}
\author{Andre Dankert}
\affiliation{\addmctwo}
\author{Raman Sankar}
\author{F. C. Chou}
\affiliation{\addtaipei}
\affiliation{\addtaipe}
\author{Paul Erhart}
\affiliation{\addphysics}
\email{erhart@chalmers.se}
\author{Saroj P. Dash}
\affiliation{\addmctwo}
\email{saroj.dash@chalmers.se}
\author{R. S. Patel}
\affiliation{\addbirla}
\email{rsp@goa.bits-pilani.ac.in}

\begin{abstract}
The Weyl semi-metal candidate MoTe$_2$ is expected to exhibit a range of exotic electronic transport properties.
It exhibits a structural phase transition near room temperature that is evident in the thermal hysteresis in resistivity and thermopower (Seebeck coefficient) as well as large spin-orbit interaction.
Here, we also document a resistivity anomaly of up to 13\%\ in the temperature window between 25 and 50\,K, which is found to be strongly anisotropic.
Based on the experimental data in conjunction with density functional theory calculations, we conjecture that the anomaly can be related to the presence of defects in the system.
These findings open opportunities for further investigations and understanding of the transport behavior in these newly discovered semi-metallic layered systems.
\end{abstract}

\maketitle

\section{Introduction}

Transition metal dichalcogenides (TMDC) have received a lot of attention due to a plethora of exciting physical phenomena and their excellent electronic, 
optical, thermal and mechanical properties \cite{keum, qi, mak, radi, splendi, ugeda, mai, Suri2017}.
TMDCs can exist in several different phases displaying a variety of electronic properties including semiconducting, metallic, superconducting, topological insulators and Weyl 
Fermionic states.
Semiconducting TMDCs possess band gaps that change from indirect to direct with the number of layers, facilitating applications such as transistors, photodetectors and electroluminescent devices \cite{Wang2012, Langouche2014, Dash2017}.
TMDCs also possess high spin-orbit coupling (SOC), which gives rise to spin polarized surface states 
in topological insulators \cite{Dankert2015} and Weyl semimetals \cite{edgeti,Soluyanov2015}.
Interestingly, the recently discovered semi-metallic phases of WTe$_2$ and MoTe$_2$ show extremely large non-saturating magnetoresistance \cite{qi, keum, mnali}, 
signatures of Weyl semimetals \cite{Deng2016, sunweyl} as well as topological Fermi arcs and surface states \cite{edgeti}, which motivates a further exploration of their fundamental properties.

Here, we investigate the Weyl semi-metal candidate MoTe$_2$ by both electronic transport experiments and density functional theory (DFT) calculations.
The basic material properties were investigated by electron, thermal and magneto-transport measurements.
We observed the structural phase transition from 1T$'$ to T$_\text{d}$ in both resistivity and Seebeck coefficient measurements, and significant spin-orbit coupling as evident from weak anti-localization signatures.
The samples exhibit a very large and strongly anisotropic resistivity anomaly in the T$_\text{d}$ phase of MoTe$_2$ that occurs between 25 and 50~K.

\section{Results and Discussions}
\subsection{Basic characterization}

MoTe$_{2}$ occurs in semiconducting or semi-metallic phases depending on the lattice structure.
In the semiconducting phase (2H) each Mo is bonded to six Te atoms in a trigonal prismatic coordination; the three Te sites above the Mo plane are located exactly on-top of the three Te sites below.
If one of the Te planes is rotated by 180 degrees one obtains the semimetallic phase (1T), which is, however, unstable toward the 1T$^{\prime}$ phase \cite{duerloo}.
The two semimetallic phases [1T$^{\prime}$ and T$_{\text{d}}$] that are in fact observed have monoclinic and orthorhombic crystal structures, respectively, and can be thought of as distorted variants of 1T.
They differ with respect to the cell angle $\beta$, which is 93.9$^{\circ}$ for 1T$^\prime$ and 90$^\circ$ for T$_{\text{d}}$.
Importantly, while 1T$^\prime$ possesses both time-reversal and inversion symmetries, T$_{\text{d}}$ does not exhibit inversion symmetry \cite{Sankar2017}.

\begin{figure}[H]
    \centering
    \includegraphics[width=15cm]{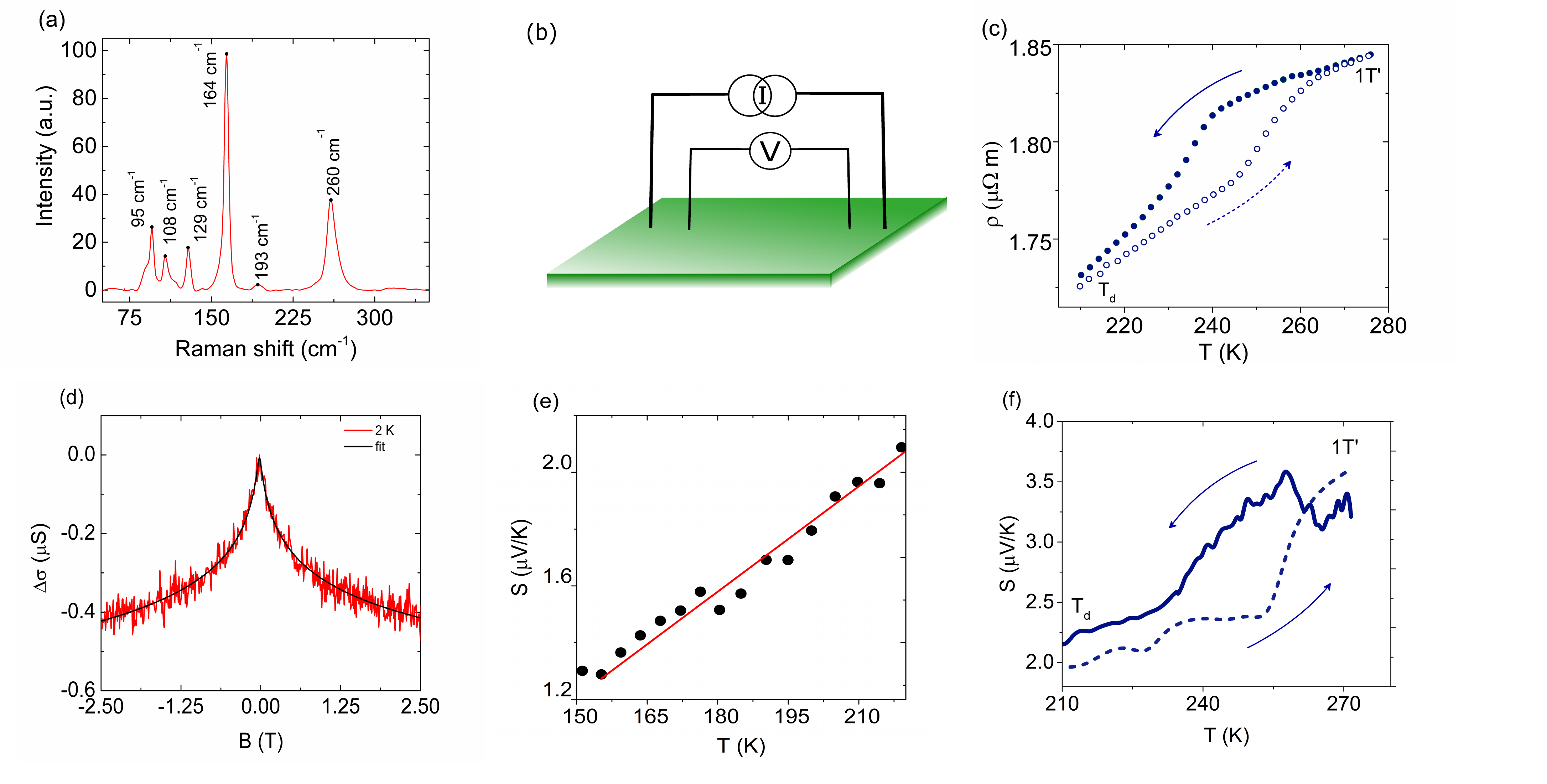}
    \caption{
        Experimental characterization.
        (a) Raman spectrum of 1T$^\prime$-MoTe$_2$ crystals measured at room temperature.
        Peaks at different wavenumbers are indicated with solid black dots.
        (b) Schematic representation of the current-in-plane (CIP) measurement configuration in the standard four probe method.
        (c) Resistivity ($\rho$) versus temperature ($T$) in the range 210--280~K in CIP configuration showing the transition from 1T$'$ to T$_{\text{d}}$ phase.
        Solid and hollow dots represent cooling and heating cycles, respectively, as indicated by arrows.
        (d) Magnetoconductance measurements: $\Delta \sigma$ versus $B$ at 2~K showing a weak anti-localization (WAL) signal.
        The red line represents the data and the black line represents the HLN fit.
        (e) Seebeck coefficient versus temperature in the range 150--220~K; black dots represent the data; the red line is the best fit curve.
        (f) Seebeck coefficient versus temperature in the range 210--280 K.
        Solid blue and dashed lines represent cooling and heating cycles, respectively.
    }
    \label{fig:characterization}
\end{figure}

MoTe$_2$ crystals were grown by the chemical vapor transport (CVT) method.
Figure \fig{fig:characterization}(a) shows the Raman spectra where the characteristic peaks corresponding to A$_g$ symmetry were observed at 108 cm$^{-1}$, 129 cm$^{-1}$, 164 cm$^{-1}$, and 260 cm$^{-1}$.
These appearance of these peaks confirms the occurrence of the 1T$^\prime$ phase at room temperature.
Furthermore a peak corresponding to B$_g$ symmetry was observed at 193 cm$^{-1}$ \cite{Oliver,Sankar2017}. 
We note that although the CVT method to grow TMDCs assures samples of high yield in less time compared to other methods, the approach results in a large number of point defects throughout the crystal.
As a result of defects and a high degree of disorder the electronic transport properties of CVT grown samples display some intriguing features \cite{PhysRevB.95.155128} and were accordingly characterized by means of resistivity, magnetoresistance, and Seebeck measurements.
Temperature dependent resistivity measurements were performed on samples of typical dimensions $5\times 2\,\text{mm}^2$, thickness $\approx\,100\,\mu\text{m}$ using the four probe technique [\fig{fig:characterization}(b)].
The resistivity as a function of temperature exhibits hysteretic behavior around 245\,K[\fig{fig:characterization}(c)], which is attributed to the structural transition from the 1T$^\prime$ phase at higher temperatures to the T$_{\text{d}}$ phase at lower temperatures \cite{Sankar2017,Zhang2016}.

Since MoTe$_2$ possesses high spin-orbit coupling, weak anti-localization (WAL) measurements were carried out as well.
In this case, application of an out-of-plane magnetic field $B_{\perp}$ breaks the time-reversal symmetry, and reduces the conductivity correction ($\Delta\sigma(B)$) as shown in \fig{fig:characterization}(d).
The conductivity reduction is a signature of WAL and can be fitted by the Hikami-Larkin-Nagaoka (HLN)  model \cite{hkn}, which yields a phase coherence length $l_{\phi}$ of $\approx\,100\,\text{nm}$.

The Seebeck coefficient increases linearly with temperature between 150 and 220\,K [\fig{fig:characterization}(e)] as expected for a metallic phase \cite{Snyder2008, reviewarxiv, ashcroft}.
The thermal hysteresis apparent in the resistivity measurements is also observed in the same temperature range in the Seebeck coefficient [\fig{fig:characterization}(f)].
As before the hysteresis is caused by the structural phase transition from the T$_{\text{d}}$ to the 1T$^\prime$ phase, and is also observed for other samples [see Fig.~S2 (b) of the Supplementary Information \cite{supp}].

\subsection{Resistivity anomaly}

\begin{figure}[H]
    \centering
    \includegraphics[width=12cm]{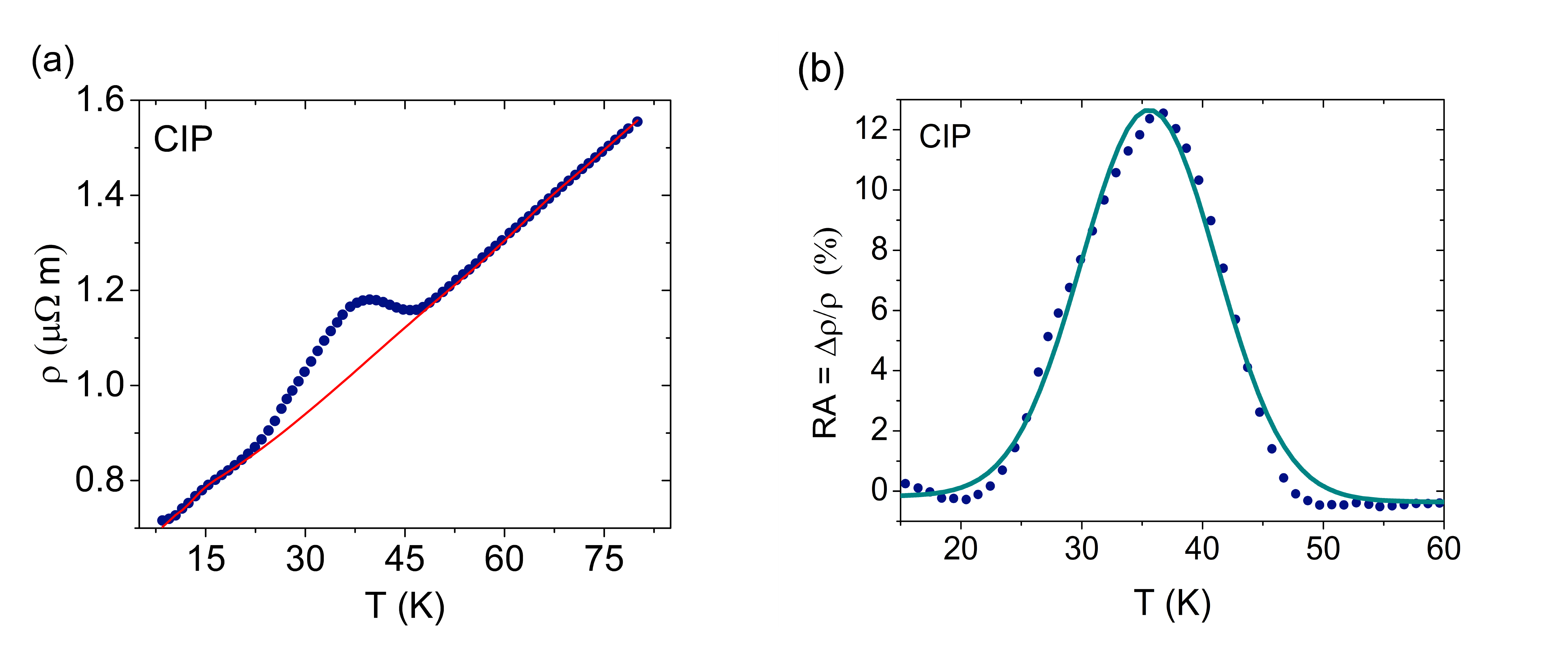}
    \caption{
        Resistivity of MoTe$_2$ in current in-plane (CIP) configuration.
        (a) Resistivity ($\rho$) versus temperature ($T$) measurements showing the resistivity anomaly in the CIP geometry.
        Measurements were performed with a constant current source of $I=100\,\text{mA}$ at a pressure of $10^{-3}\,\text{mbar}$.
        The red line represents the background as described by a polynomial fit. 
        (b) Resistivity anomaly ($\frac{\Delta\rho}{\rho}$) versus temperature, in the range 20--60\,K, plotted after background subtraction.
        Data points are shown by blue dots while the cyan line represents a Gaussian fit.
    }
    \label{fig:anomaly-cip}
\end{figure}

Having established that the basic characteristics of our samples in the temperature window between approximately 150\,K to room temperature match those expected for the known semi-metallic phases of MoTe$_2$, we can now focus on the low-temperature behavior.
Here, the temperature dependent resistivity obtained at a pressure of $10^{-3}\,\text{mbar}$ reveals a resistivity anomaly (RA) that sets in around a particular temperature of T$_{\text{RA}} \approx$36~K [\fig{fig:anomaly-cip}].
The resistivity anomaly is defined as
\begin{align}
    \text{RA (\%)}
    &= \frac{\rho_{\text{raw data}} - \rho_{\text{background}}}{\rho_{\text{background}}} \times 100 \%
    = \frac{\Delta \rho}{\rho} \times 100 \%
\end{align}
In the present case, we find a RA of up to 13\,\%, which is best described by a Gaussian fit.
Such anomalies in TMDCs have been found to be sensitive to various factors including, e.g., impurities, doping, and pressure \cite{zocco, ritschel}.
Our measurements have been repeated on different samples and reproduced consistently (see Figs.~S3 (a,b) of the Supplementary Information \cite{supp}).
To confirm that the signature is purely from the sample, a control experiment was also performed on a Cu wire in similar experimental conditions, which showed normal metallic behavior [Fig.~S3 (c)].
Given the growth method, which as discussed above tends to be associated with a relatively high defect density and the WAL analysis, We tentatively attribute these anomalies to defects incorporated during synthesis.

\begin{figure}[H]
\centering
    \includegraphics[width=15cm]{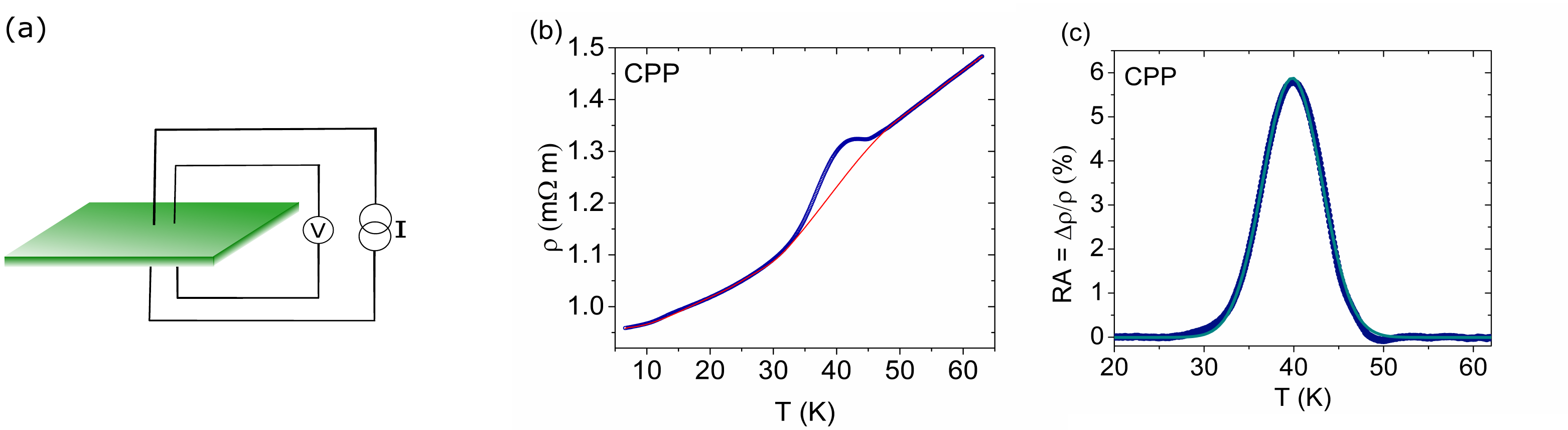}
    \caption{
        Resistivity anomaly of MoTe$_2$ in current-perpendicular-to-plane (CPP) configuration.
        (a) Schematic illustration of CPP measurement in four probe configuration.
        (b) Resistivity ($\rho$) versus temperature ($T$) measurements showing the resistance anomaly in the CPP measurement geometry. The red line represents the polynomial curve used for background subtraction.
        (c) Resitivity anomaly ($\frac{\Delta\rho}{\rho}$) versus temperature, in the range 20--60~K, plotted after background subtraction. Data points are shown by blue dots while the cyan line represents a Gaussian fit.
    }
    \label{fig:anomaly-cpp}
\end{figure}

In MoTe$_2$, the intra-layer bonding is strong and covalent, whereas the inter-layer bonding is due to weak van-der-Waals forces, leading to strong anisotropy in many properties.
Measurements were therefore also performed in current-perpendicular-to-plane (CPP) configuration, 
where the current
flows perpendicular to the sample planes [\fig{fig:anomaly-cpp}(a)].
The thermal hysteresis (corresponding to the 1T$'$-T$_\text{d}$ structural transition close to room temperature)  observed in CPP mode was centered at 254~K.
The degree of anisotropy estimated from the comparison of the resistivity measured in CIP and CPP modes is $\approx$ 10$^{3}$.
In the CPP measurement configuration $T_{\text{RA}}$ is centered at 40~K [\fig{fig:anomaly-cpp}(b)], which is close to the temperature of 36~K obtained in CIP geometry.
The RA obtained in CPP mode is, however, 6\,\%\ and thus about half of that observed in CIP mode (13.1\%) [\fig{fig:anomaly-cpp}(c)].
The reduction of the resistivity anomaly in CPP mode compared to CIP mode may be attributed to the high degree of structural anisotropy.
In particular, since the weak van-der-Waals gap inter-layers bonding is associated with soft phonon modes, one can expect the electron-phonon coupling strength to be very anisotropic.

\begin{figure}[H]
    \centering
    \includegraphics[width=17cm]{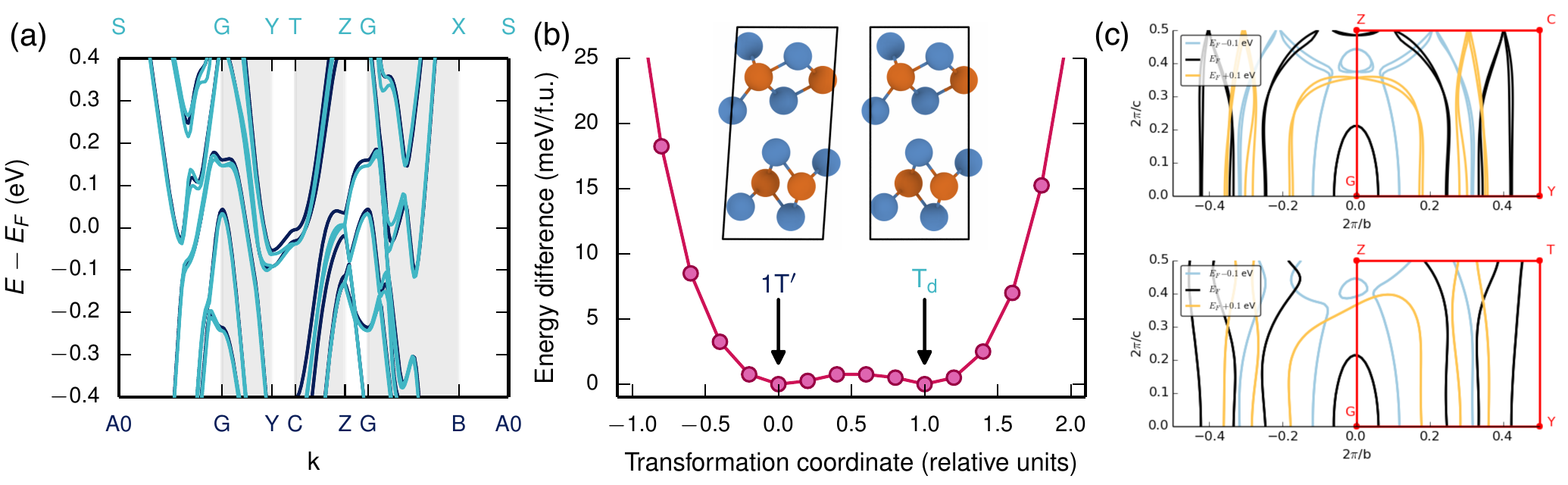}
    \caption{
        (a) Electronic band structure of 1T$^{\prime}$-MoTe$_2$ (blue) and T$_\text{d}$-MoTe$_2$ (black lines).
        (b) Energy landscape for the transition from the 1T$^{\prime}$ phase to T$_{\text d}$.
        (c) Projected Fermi surface on the $k_y-k_z$ plane.
        Notice the open orbits that are present in the 1T$^{\prime}$ phase in contrast to the pinching at the zone boundary that occurs in the T$_{\text d}$ phase.}
    \label{fig:dft}
\end{figure}

In order to gain further understanding of the experimental results, density functional theory (DFT) calculations were performed using the project augmented wave (PAW) method \cite{Blo94, KreJou99} as implemented in the Vienna ab-initio simulation package \cite{KreFur96a}.
Structural relaxation and total energy calculations were carried out using the vdW-DF-cx exchange-correlation functional \cite{DioRydSch04, BerHyl14}, which has been previously shown to provide an excellent description of the lattice structure of semiconducting TMDCs with van-der-Waals bonding \cite{LinErh16}.
While spin-orbit coupling (SOC) is important given the large mass of Te, the spin-polarized version of vdW-DF-cx has not yet been verified for such calculations yet.
The electronic structure including SOC effects was therefore computed using the PBE fucntional \cite{PerBurErn96} on the vdW-DF-cx relaxed structures.
The structures were relaxed until the maximum force in the system was less than 5~meV/{\AA}.
The plane wave cutoff energy was set to 400~eV and the Brillouin zone was sampled suing a $24\times12\times6$ Monkhorst-Pack grid.

While the calculated in-plane lattice parameters are in very good agreement with experiment, the out-of-plane lattice parameter in the metallic phases of MoTe$_2$ is underestimated by $2.5\%$ compared to experimental data \cite{Sankar2017}.
In this context, it is interesting to note that in the case of WSe$_2$ a similar difference for the out-of-plane lattice constant was observed between conventionally and turbostratically grown crystals of the 2H phase \cite{NguBerLin10, ErhHylLin15}, which could be attributed to the larger density of stacking faults in the latter material.
Since the vdW-DF-cx method otherwise reproduces the results for the ideal crystal structures of TMDCs very well, the discrepancy could be another telltale sign of the rather higher defect density in the present samples that was already alluded to above.
A comprehensive comparison of the calculated lattice parameters can be found in the Supplementary Material.

The electronic band structures of bulk 1T$^\prime$ and T$_{\text{d}}$ MoTe$_2$ show a multi-valley structure with many pockets and band inversions.
Generally, they are very similar [Fig.~\ref{fig:dft}~(a)] with noticeable differences only along the Y-C (Y-T) and C-Z (T-Z) paths in the Brillouin zone of T$_\text{d}$ (1T$^{\prime}$).
According to our calculations, the ideal 1T$^{\prime}$ and T$_{\text d}$ structures are energetically practically degenerate and separated by a very small transition barrier of about 1~meV/f.u. [Fig.~\ref{fig:dft}~(b)].
This extremely soft landscape suggests that these phases can be very sensitive to thermal perturbations and defects. The most striking difference in the electronic band structure is the pinching of the energy bands close to the Fermi energy at the zone boundary near $Z$ in the T$_{\text d}$ phase [Fig.~\ref{fig:dft}~(c)], a feature that is absent in the case of 1T$^{\prime}$.
Qualitatively, this means that the orbits in T$_{\text d}$ are closed, whereas the obits are open in 1T$^{\prime}$.
We note, however, that the precise location of (and number of) the Weyl points in the T$_{\text d}$ phase is known to be very sensitive to the structural parameters \cite{TamWuCuc16}, and even small changes in the lattice parameter (due to thermal expansion or defects) might change the Fermi surface considerably and thus affect electrical transport in the material.

\section{Conclusions}
To summarize, we have carried out transport measurements on semi-metallic MoTe$_2$ in various temperature ranges.   MoTe$_2$ showed semi-metallic properties with high spin-orbit coupling, structural phase transitions from T$_d$ to 1T$^{\prime}$  near to room temperature in the temperature dependent resistivity (in both CIP and CPP measurement geometries) and Seebeck coefficient measurements. We observed a pronounced resistivity anomaly up to 13\% at low temperature with a strong anisotropy of $\approx$ 10$^3$ (three orders of magnitude) between the CIP and CPP measurement configuration. The resistivity anomaly can be attributed to point defects created during the synthesis of MoTe$_2$ by chemical vapor transport process. The DFT calculations performed suggest that the resistivity anomaly is not present in the pristine crystal, which further supports the attribution to defects. This work forges a path for further investigation of metastable state leading to reorganization in the electron density and its relation to chiral symmetry breaking in various Weyl semi-metal candidates.

\section{Acknowledgements} DS thanks Department of Science and Technology (DST), Government of India for a PhD fellowship through the DST-INSPIRE program (No. DST/INSPIRE Fellowship/2013/742). RSP thanks the DST, Government of India for financial support through the Nanomission program (No. SR/NM/NS-1002/2010 (G)) and SERB grant (No. EMR/2016/003318). SPD thanks financial supports from the European Union Graphene Flagship (No. 604391), a FlagEra project (VR No. 2015-06813), and the Swedish Research Council (No. 2016-03658). CL, SKS, and PE acknowledge support from the Knut and Alice Wallenberg Foundation.\\~\\
\noindent \textbf{Corresponding Authors:}\\
R. S. Patel: rsp@goa.bits-pilani.ac.in; Saroj P. Dash: saroj.dash@chalmers.se; Paul Erhart: erhart@chalmers.se

\end{document}